\documentclass[a4paper]{article}

\usepackage{
  latexsym
  ,amssymb
  ,amsmath
  ,mathptm
  }
\newcommand{\PCPHook}{%
  \mathbin{%
    \hbox{%
      \!\vrule height 0.4pt width 5.0pt%
      }%
    \hbox{\vrule height 6.0pt width 0.4pt}\;
    }
  }
\newcommand{\PCPsstapeln}[2]{%
  \!\raisebox{0.7ex}{$\scriptscriptstyle\stackrel{#1}{#2}$}}
\newcommand{\PCPistapeln}[2]{%
  \!\raisebox{-0.3ex}{$\scriptscriptstyle\stackrel{#1}{#2}$}}
\newcommand{\PCPstapeln}[2]{{\scriptstyle\genfrac{}{}{0pt}{}{#1}{#2}}}

\def\PCPqed{\hspace*{\fill}\ensuremath{\Box}\\}
\newcommand{\PCPdbyd}[1]{\ensuremath{%
   \frac{d}{d{#1}}\PCPatpoint{#1=0}}}
\newcommand{\PCPatpoint}[1]{\,\rule[-1.55ex]{0.1ex}{4.6ex}\,
        \raisebox{-1.55ex}{\ensuremath{\scriptscriptstyle #1}}\,}
\newcommand{\PCPrestrict}[1]{%
        \raisebox{0ex}{\ensuremath{\restriction}}
        \hspace*{-0.52ex}\rule[-1.4ex]{0.09ex}{2.5ex}\,
        \raisebox{-1.4ex}{\ensuremath{\scriptscriptstyle #1}}}
\newcommand{\PCPdel}{\ensuremath{\partial}}

\IfFileExists{cmcal.sty}{\usepackage{cmcal}}
                        {\newcommand{\cmcal}[1]{\mathcal{##1}}}
\def\PCPcal#1{%
     \ensuremath{\cmcal{#1}}}
\IfFileExists{bbm.sty}{\usepackage{bbm}}
                        {\newcommand{\mathbbm}[1]{\mathbb{##1}}}

\def\PCPbbm#1{%
     \ensuremath{\mathbbm{#1}}}
\def\PCPfrak#1{%
     \ensuremath{\mathfrak{#1}}}
\newtheorem{elemma}{Lemma}[section]
\newtheorem{prop}{Proposition}[section]
\newcommand{\PCPbV}{\ensuremath{\bar{\PCPcal V}} }
\newcommand{\PCPbVast}{\ensuremath{\bar{\PCPcal V}^\ast} }
\newcommand{\PCPbG}{\ensuremath{\bar{\Gamma}} }
\newcommand{\PCPbGast}{\ensuremath{\bar{\Gamma}^\ast} }
\newcommand{\PCPbv}{{\ensuremath{\bar{v}}} }
\newcommand{\PCPVV}{\ensuremath{\PCPfrak V\PCPcal V} }
\newcommand{\PCPJV}{\ensuremath{\PCPfrak J^1\PCPcal V} }
\newcommand{\PCPJVast}{\ensuremath{\left(\PCPfrak J^1\PCPcal V\right)^\ast} }
\newcommand{\PCPTJVast}{\ensuremath{T\left(\PCPfrak J^1\PCPcal V\right)^\ast} }
\newcommand{\PCPxx}[1]{\raisebox{1pt}{\ensuremath{\stackrel{#1}{X}}}}
\newcommand{\PCPF}[1]{\raisebox{1pt}{\ensuremath{\stackrel{#1}{F}}}} 
\newcommand{\PCPG}[1]{\raisebox{1pt}{\ensuremath{\stackrel{#1}{G}}}} 
\newcommand{\PCPHH}[1]{\raisebox{1pt}{\ensuremath{\stackrel{#1}{H}}}} 
\newcommand{\PCPoV}{\ensuremath{\Omega^{(2,n)}}}
\begin{document}

\title{%
  \begin{flushright}
    \small Freiburg THEP-99/14\\
    To appear in Rep. on Math. Phys. {\bf 47}, 2001
  \end{flushright}
  \vspace{3ex}
  A vertical exterior derivative in multisymplectic geometry
  and a graded Poisson bracket for nontrivial geometries
  }

\author{%
  Cornelius Pauf\/ler\thanks{%
    e-mail: pauf\/ler@physik.uni-freiburg.de%
    }\\
  Fakult\"at f\"ur Physik\\ 
  Albert-Ludwigs-Universit\"at Freiburg im Breisgau\\ 
  Hermann-Herder-Stra\ss e 3\\ 
  D 79104 Freiburg i. Br.\\
  Germany
  }

\date{February 9, 2000}

\maketitle

\begin{abstract}
  A vertical exterior derivative is constructed that is needed for a
  graded Poisson structure on multisymplectic mani\-folds over
  nontrivial vector bundles. In addition, the properties of the
  Poisson bracket are proved and first examples are discussed.
\end{abstract}
\section{Introduction}

In \cite{GoIsMa98} a geometrical framework to handle field theories over
manifolds in a finite
dimensional geometry is proposed. 
This mathematical setting appears under the name
multisymplectic geometry, De Donder-Weyl theory, Hamilton-Cartan formalism, and
covariant field theory in the literature 
(\cite{GoSt73,Di91,GiMaSa99}, further \cite{Gue87,Gue91}). 
The basic idea is to treat the space coordinates of a given field theory as
additional evolution parameters. Thus, there is a finite number of variables
(the field and its first derivatives) that evolve in space-time rather than a
curve in an infinite-dimensional vector space of field configurations. 
As shown
in \cite{Trau67,Ru66} one can incorporate the field equations and the Noether
theorem \cite{Noe18b} in that formulation, but in order to find a corresponding
quantum field theory -- at least in the sense of a formal deformation
\cite{BFFLS78,Dit90} -- one  has to formulate the dynamics of the classical
theory in terms of Poisson brackets first.

Kanatchikov  (\cite{Kan98,Kan99}) has constructed such a bracket 
for trivial vector bundles over orientable manifolds. In the
nontrivial case the 
used ``vertical exterior derivative'' which plays a central r\^ole in
the construction is  
not globally defined 
(the resulting bracket, however, does not depend on the coordinate
system used). 
What is needed is a derivative in vertical directions that 
in particular has square zero. A first guess would be to use a connection and
take an expression like 
$dv^A\wedge\nabla_A$ with $\nabla$ being a covariant derivative and
$dv^A$ being  
vertical. 
The condition that its square gives zero is then equivalent to the flatness of
$\nabla$ along fibres. As the fibres under consideration are vector 
spaces one would indeed expect that it is possible to construct such a
covariant derivative. This construction constitutes the main part of
this paper. 

The remaining part of this article is organised as follows. In the
first section a short overview over the multisymplectic approach is
given. Then, with the help of a covariant derivative that is flat
along the fibres of phase space, the already mentioned vertical
exterior derivative is constructed and discussed. Then the Poisson
structure is given and the defining properties are proved. Finally,
mechanics as the case of a trivial (vector) bundle over a
one-dimensional base manifold (i.e., the time axis \PCPbbm R) is
recovered and the scalar field case is considered.

The appendix contains some well known facts about connections viewed
as sections of jet bundles and the construction of the already
mentioned covariant derivative on the multisymplectic phase space.

\section{From variational principles to multisymplectic geometry}

In field theory, solutions of the field equations are stationary points of the
action functional
\[
L[\varphi]=\int_{\PCPcal M}\,\PCPcal
L\left(\varphi(x),\nabla\varphi(x)\right)\,d^{n+1}x,
\]
where \PCPcal M is some $(n+1)$-dimensional parameter space (e.g. space-time),
$\nabla\varphi$ is the gradient of the field $\varphi$ and \PCPcal L is the
Lagrange density.

In general, $\varphi$ is a section of a vector bundle 
$\pi:\PCPcal V\rightarrow\PCPcal M$. 
$T\varphi:T\PCPcal M\rightarrow T\PCPcal V$ 
fulfils 
$T\pi\circ T\varphi = T\textrm{id}_{\PCPcal M}$ 
and thus\footnote{%
  Usually, the first jet bundle of a vector bundle 
  $(\PCPcal M,\pi,\PCPcal V)$ is defined to be the set of all
  equivalence classes at a point of \PCPcal M of local sections, where
  equivalence means equal function value and first derivatives. But
  this can be viewed as a tangent map from $T\PCPcal M$ to 
  $T\PCPcal V$ having the stated property. Further, such a tangent map
  defines how to (horizontally) lift $T\PCPcal M$ at every point of
  \PCPcal V, which is equivalent to having a connection. Hence, a
  connection defines a map 
  $\PCPcal V\rightarrow\PCPcal J^1\PCPcal V$, 
  which turns the affine bundle $\PCPcal J^1\PCPcal V$ 
  into a vector bundle over \PCPcal V.
}%
defines an element of $\PCPfrak J^1\PCPcal V$, the
first jet bundle of \PCPcal V (\cite{Sau89,KMS93}). Using a linear connection
\[
   \Gamma:\PCPcal V\rightarrow\PCPfrak J^1\PCPcal V
\]
we obtain an isomorphism 
\[
   i_\Gamma:\left(\PCPfrak J^1\PCPcal V\right)_v\rightarrow
   \left(\PCPcal V\otimes T^\ast\PCPcal M\right)_{\pi(v)}
\]
for all $v$ in \PCPcal V, where in addition we have used 
$\left(\PCPfrak V\PCPcal V\right)_v\cong\PCPcal V_{\pi(v)}$ 
for vector bundles \PCPcal V and their vertical
tangent bundles $\PCPfrak V\PCPcal V$. In particular,
we find
\begin{equation}\label{specForm}
  i_\Gamma\circ T_x\varphi\circ\xi(x)=\nabla_{\xi}\varphi(x),
\end{equation}
for $\nabla$ denoting the covariant derivative corresponding to $\Gamma$ and 
$\xi$ being a tangent vector on \PCPcal M. This will be needed in section
4.

Now the Lagrange density can be interpreted as a mapping 
\[
   \PCPcal L:\PCPfrak J^1\PCPcal V\rightarrow\Lambda^{n+1}T^\ast\PCPcal M,
   \quad
   L[\varphi]=\int_{\PCPcal M}\,\PCPcal L\circ j^1\varphi,
\]
where 
   $j^1\varphi(x)
      = T_x\phi\in\left(\PCPfrak J^1\PCPcal V\right)_{\varphi(x)}$ 
is the first jet prolongation of $\varphi\in\Gamma(\PCPcal V)$.
Stationary points of $L$ correspond to solutions of the Euler-Lagrange
equations, which in local coordinates\footnote{%
  When working in local coordinates of \PCPJV we will use the
  following convention. Small Latin indices sum over the base manifold
  directions, that is $i,j,k$ run from $1$ to $n+1$ if not specified
  otherwise. Capital Latin characters as $A,B,C,D$ run from $1$ to $N$
  which is the dimension of a fibre of \PCPcal V.
}%
$(x^i,v^A,v^A_i)$ of $\PCPfrak J^1\PCPcal V$ read (cf. \cite{Ru66})
\begin{equation}\label{ELG}
  \frac{\PCPdel\PCPcal L}{\PCPdel v^A}\circ j^1\varphi
  -\frac{\PCPdel}{\PCPdel x^i}
  \left(
    \frac{\PCPdel\PCPcal L}{\PCPdel v^A_i}\circ j^1\varphi
  \right)
  =0.
\end{equation}
Now we want to formulate the theory on what we shall call phase space. Since
$\PCPfrak J^1\PCPcal V$ is not a vector bundle but an affine bundle,
one chooses the dual $\left(\PCPfrak J^1\PCPcal V\right)^\ast$ to be
the bundle of affine mappings from $\PCPfrak J^1\PCPcal V$ 
to $\Lambda^{n+1}T^\ast\PCPcal M$. Thus, coordinates $(x^i,v^A,v^A_i)$
on $\PCPfrak J^1\PCPcal V$ induce coordinates $(x^i,v^A,p,p^i_A)$ 
on $\left(\PCPfrak J^1\PCPcal V\right)^\ast$. One can show (see
\cite{GoIsMa98}, ch. 2B) that $\PCPJVast$, being a vector bundle over
\PCPcal V (it inherits a vector space structure from the target space
$\Lambda^{n+1}T^\ast\PCPcal M$), is canonically isomorphic to
$\PCPcal Z\subset\Lambda^{n+1} T^\ast \PCPcal V$, where
\[
   \PCPcal Z_v=\{z\in\Lambda^{n+1}T^\ast\PCPcal V_v | i_V\,i_W\,z=0\;
   \forall V,W \in\left(\PCPfrak V\PCPcal V\right)_v\},
   \PCPcal Z=\bigcup_{v\in\PCPcal V}\PCPcal Z_v.
\]
Furthermore, on $\Lambda^{n+1}T^\ast\PCPcal V$ there is a canonical
$(n+1)$-form $\Theta_\Lambda$, defined by
\[
   \Theta_\Lambda(z)(u_1,\ldots,u_{n+1})
   =
   z(T\pi_{\PCPcal V\Lambda}u_1,\ldots,T\pi_{\PCPcal V\Lambda}u_{n+1}),
\]
where $z\in\Lambda^{n+1}T^\ast\PCPcal V$, 
$u_1,\ldots,u_{n+1}\in T_z\Lambda^{n+1}T^\ast\PCPcal V$, 
$\pi_{\PCPcal V\Lambda}:\Lambda^{n+1}T^\ast\PCPcal V\rightarrow\PCPcal V$.
Using the embedding 
$i_{\Lambda\PCPcal Z}:
    \PCPcal Z\rightarrow\Lambda^{n+1}T^\ast\PCPcal V$, 
we obtain an $(n+1)$-form on \PCPcal Z,
\begin{equation}\label{can.n+1.form}
  \Theta=i_{\Lambda\PCPcal Z}^\ast\Theta_\Lambda,
\end{equation}
which will be called canonical $(n+1)$-form thereafter.
There is a canonical $(n+2)$-form $\Omega$ on \PCPcal Z, too,
\[
   \Omega=-d\Theta.
\]
Using coordinates $(x^i,v^A,p,p^i_A)$, one finds
\[
   \Theta=p^i_A\,dv^A\wedge\left(\PCPdel_{x^i}\PCPHook d^{n+1}x\right)
           +p\,d^{n+1} x,
   \quad
   \Omega=
           dv^A\wedge dp^i_A\wedge
           \left(\PCPdel_{x^i}\PCPHook d^{n+1}x\right)
           -dp\wedge d^{n+1} x,
\]
where $d^{n+1}x=dx^1\wedge\cdots\wedge dx^{n+1}$. Now we are in the position to
reformulate (\ref{ELG}). As a first step we define a covariant
Legendre transform for $\PCPcal L$: 
\begin{equation}\label{Leg-Trf}
  \begin{split}
    \PCPbbm F\PCPcal L:\PCPfrak J^1\PCPcal V\ni\gamma
    &\mapsto
    \PCPbbm F\PCPcal L(\gamma)
    \in\left(\PCPfrak J^1\PCPcal V\right)^\ast\cong\PCPcal Z,
    \\
    \PCPbbm F\PCPcal L(\gamma):
    \PCPfrak J^1\PCPcal V\ni\gamma'
    &\mapsto
    \PCPcal L(\gamma)+
    \PCPdbyd\epsilon\PCPcal L\left(\gamma+\epsilon(\gamma'-\gamma)\right)
    \in\Lambda^{n+1}T^\ast\PCPcal M.
  \end{split}
\end{equation}
In coordinates as above it takes the form
\begin{equation}\label{Leg-Koord}
  \PCPcal L=L(x^i,v^A,v^A_i)\,d^{n+1}x,
  \quad
  p_A^i=\frac{\PCPdel L}{\PCPdel v^A_i},
  \quad
  p=L-\frac{\PCPdel L}{\PCPdel v^A_i}v^A_i.
\end{equation}
Using $\PCPbbm F\PCPcal L$ we can pull back the canonical $(n+1)$-form $\Omega$
to obtain the so-called Cartan form $\Theta_{\PCPcal L}$,
\[
   \Theta_{\PCPcal L}=\left(\PCPbbm F\PCPcal L\right)^\ast\Theta.
\]
One can show (\cite{GoIsMa98}, theorem 3.1) that the Euler-Lagrange
equations (\ref{ELG}) are equivalent to
\[
   \left(j^1\varphi\right)^\ast(i_W\Omega_{\PCPcal L})=0\quad
   \forall W\in T\PCPfrak J^1\PCPcal V,
\]
where 
\[
   \Omega_{\PCPcal L}=-d\Theta_{\PCPcal L}
   =
   \left(\PCPbbm F\PCPcal L\right)^\ast\Omega.
\]

\section{A vertical exterior derivative}

Let us denote the multisymplectic phase space \PCPJVast by \PCPcal P
to simplify notation. In what follows we will need a mapping that is
in some sense the vertical part of the exterior derivative on \PCPcal
P. In particular, it must have square zero. Whereas the derivation
along fibres of $\PCPcal P\rightarrow\PCPcal M$ can be defined without
additional data, the space of vertical forms as a subspace of
arbitrary forms cannot\footnote{%
  One can, however, define the space of vertical forms canonically,
  but in what follows we need the wedge product of a vertical form and
  an arbitrary one. For this, one needs an embedding of vertical forms
  in the space of forms, which in turn requires the use of a
  connection.
}.%
This is due to the fact that one needs to specify what is {\sl not}
vertical if one is looking for the dual of vertical vectors. For this,
one needs a connection in the bundle \PCPcal P over \PCPcal M. This is
dealt with in appendix A. With the help of this connection we can
split $T_p\PCPcal P$ into horizontal and vertical components for each
point $p$ of \PCPcal P. In local coordinates\footnote{%
  When working in coordinates of \PCPcal P, we will use the following
  convention which is similar to the one for coordinates on
  \PCPJV. Small Latin indices sum over the base manifold directions,
  that is $i,j,k$ run from $1$ to $n+1$ if not specified
  otherwise. Capital Latin characters as $A,B,C,D$ run from $1$ to $N$
  which is the dimension of a fibre of \PCPcal V. Small  Greek indices
  can be both base manifold and \PCPcal V-fibre and dual jet bundle
  indices, i.e. $\rho,\sigma,\tau=i,A,\PCPstapeln iA$. Finally,
  capital letters from $M$ onwards stand for both $A,B\ldots$ and
  $\PCPstapeln iA,\PCPstapeln jB,\ldots$.
}%
$(x^i,v^A,p^i_A,p)$ we have a basis
$(\PCPfrak e^{\ast \alpha}_{(p)},\PCPfrak e)$, $\alpha=i,A,\PCPstapeln iA$ 
of $T^\ast_p\PCPcal P$ that is dual to a basis 
$(\PCPfrak e_{\alpha}(p),\PCPfrak e)$ of $T_p\PCPcal P$. The detailed
definition of the latter is explained in the appendix. In coordinates as above,
\begin{equation}\label{split}
  \PCPfrak e^{\ast i}_{(p)}=dx^i,\quad
  \PCPfrak e^{\ast A}_{(p)}=dv^A+\Gamma_{iB}^A(\pi(p))\,v^B\,d x^i,\quad 
  \PCPfrak e^{\ast \PCPistapeln iA}_{(p)}=dp^{\PCPstapeln iA}
  +(\Lambda^i_{kj}\delta_A^B-\Gamma_{kA}^B\delta_j^i)\,p^{\PCPstapeln kB}\,dx^j
  ,\quad \PCPfrak e^\ast_{(p)}=dp.
\end{equation}
Using the duality between $T\PCPcal P$ and $T^\ast\PCPcal P$, we
obtain a covariant derivative $D^\ast$ on $T\PCPcal P$, in particular
\begin{equation*}
  \begin{array}{rccclrcccl}
    \left({D^\ast}_{\PCPfrak e_M}\PCPfrak e^{\ast N}\right)
    (\PCPfrak e_\rho)(p)
    &=&
    -\PCPfrak e^{\ast N}
    \left({D}_{\PCPfrak e_M}\PCPfrak e_\rho\right)(p)
    &=&0,
    &\quad
    \left({D^\ast}_{\PCPfrak e_M}\PCPfrak e^{\ast i}\right)
    (\PCPfrak e_\rho)(p)
    &=&-\PCPfrak e^{\ast i}
    \left({D}_{\PCPfrak e_M}\PCPfrak e_\rho\right)(p)&=&0
  \end{array}
\end{equation*}
for all fibre indices $M,N=A,\PCPstapeln iA$ and all indices $\rho$.
Thus for every 
$\alpha{(p)}
    =\frac 1{l!}{\alpha_{\rho_1\cdots\rho_l}}_{(p)}
       \PCPfrak e^{\ast\rho_1}_{(p)}\wedge\cdots\wedge
       \PCPfrak e^{\ast\rho_l}_{(p)}\in \Omega^l\PCPcal P
    =\Gamma(\Lambda^lT^\ast\PCPcal P)
$ 
the mapping\footnote{%
  This mapping is a globally defined version of the vertical
  differential used by Kanatchikov in \cite{Kan98,Kan99}.
}%
\[
   d^V
   =\left(\PCPfrak e^{\ast M}_{(p)}\wedge{D^\ast}_{\PCPfrak e_M}\right):
   \Omega^l\PCPcal P\rightarrow\Omega^{l+1}\PCPcal P
\]
fulfils ($M,N=A,\PCPstapeln iA$ for $i=1,\ldots,n$, $A=1,\ldots,N$,
$\rho_l=i,A,\PCPstapeln iA$) 
\[
\begin{split}
  \left(d^V\right)^2\alpha{(p)}
  &=
  \left(d^V\right)^2
  \frac 1{l!}
  {\alpha_{\rho_1\cdots\rho_l}}_{(p)}
  \PCPfrak e^{\ast\rho_1}_{(p)}\wedge\cdots\wedge
  \PCPfrak e^{\ast\rho_l}_{(p)}
\\
  &=
  \left(\PCPfrak e^{\ast M}_{(p)}\wedge{D^\ast}_{\PCPfrak e_M}\right)
  \left(\PCPfrak e^{\ast N}_{(p)}
    \wedge{D^\ast}_{\PCPfrak e_N}
  \right)
  \frac 1{l!}{\alpha_{\rho_1\cdots\rho_l}}_{(p)}
  \PCPfrak e^{\ast\rho_1}_{(p)}\wedge\cdots\wedge
  \PCPfrak e^{\ast\rho_l}_{(p)}
\\
  &=
  \frac 1{l!}\left(
    \PCPfrak e^{\ast M}_{(p)}\wedge{D^\ast}_{\PCPfrak e_M}
  \right)
  \left(\PCPfrak e_N\,\alpha_{\rho_1\cdots\rho_l}\right)_{(p)}
  \PCPfrak e^{\ast N}_{(p)}\wedge 
  \PCPfrak e^{\ast\rho_1}_{(p)}\wedge\cdots\wedge
  \PCPfrak e^{\ast\rho_l}_{(p)}
\\
  &\quad
  +\frac 1{l!}\left(
    \PCPfrak e^{\ast M}_{(p)}\wedge{D^\ast}_{\PCPfrak e_M}
  \right)
  \sum_{k=1}^l {\alpha_{\rho_1\cdots\rho_l}}_{(p)}
  \PCPfrak e^{\ast N}_{(p)}\wedge \PCPfrak e^{\ast\rho_1}_{(p)}\wedge\cdots
  \wedge
  \underbrace{{D^\ast}_{e_N}\PCPfrak e^{\ast \rho_k}}_{=0}
  \wedge\cdots\wedge \PCPfrak e^{\ast\rho_l}_{(p)}
\\
  &=
  \frac 1{l!}\left(
    \PCPfrak e_M\,\PCPfrak e_N\,\alpha_{\rho_1\cdots\rho_l}
  \right)_{(p)}
  \PCPfrak e^{\ast M}_{(p)}\wedge 
  \PCPfrak e^{\ast N}_{(p)}\wedge 
  \PCPfrak e^{\ast\rho_1}_{(p)}\wedge\cdots\wedge
  \PCPfrak e^{\ast\rho_l}_{(p)}
\\[1ex]
  &\quad
  +\frac 1{l!}\left(
    \PCPfrak e_N\,\alpha_{\rho_1\cdots\rho_l}
  \right)_{(p)}
  \PCPfrak e^{\ast M}_{(p)}\wedge 
  \underbrace{{D^\ast}_{\PCPfrak e_M}\PCPfrak e^{\ast N}}_{=0}
  \wedge \PCPfrak e^{\ast\rho_1}_{(p)}
  \wedge\cdots\wedge \PCPfrak e^{\ast\rho_l}_{(p)}
\\
  &\quad+
  \frac 1{l!}\sum_{k=1}^l
  \left(\PCPfrak e_N\,\alpha_{\rho_1\cdots\rho_l}\right)_{(p)}
  \PCPfrak e^{\ast M}_{(p)}\wedge \PCPfrak e^{\ast N}_{(p)}\wedge 
  \PCPfrak e^{\ast\rho_1}_{(p)}\wedge\dots
  \wedge
  \underbrace{{D^\ast}_{e_N}\PCPfrak e^{\ast \rho_k}}_{=0}
  \wedge\dots\wedge \PCPfrak e^{\ast\rho_l}_{(p)}
\\
  &=\frac 1{2l!}\left(
    [\PCPfrak e_M,\PCPfrak e_N]\,\alpha_{\rho_1\cdots\rho_l}
  \right)_{(p)}
  \PCPfrak e^{\ast M}_{(p)}\wedge 
  \PCPfrak e^{\ast N}_{(p)}\wedge 
  \PCPfrak e^{\ast\rho_1}_{(p)}\wedge\cdots\wedge
  \PCPfrak e^{\ast\rho_l}_{(p)}
\\
  &=0,
\end{split}
\]
that is, $\left(d^V\right)^2=0$. This justifies the name vertical
exterior derivative. 

\subsection{Poincar\'e lemma for $d^V$}

\begin{elemma}[Poincar\'e lemma for $d^V$]\label{PoinLemma}
  Let $\alpha\in\Omega^r\PCPcal P$ with $d^V\alpha=0$. Then for every
  $p\in\PCPcal P$ there exists a neighbourhood $U_p$ and a $(r-1)$-Form
  $\beta$ such that $\alpha\PCPrestrict{\PCPcal U_p}=d^V\,\beta$.
\end{elemma}
{\sc Proof:}
As fibres of $\PCPcal P\rightarrow\PCPcal M$ are contractible and
$d^V$, restricted to such a fibre, acts like the exterior derivative,
this is a consequence of the Poincar\'e lemma. In detail, let
$m=\pi(p)$ and \PCPcal U be a neighbourhood of $m$ such that 
$\PCPcal P\PCPrestrict{\PCPcal U}$ is trivial. Now let 
$\PCPcal U_p=\pi^{-1}(\PCPcal U)$. On $\PCPcal U_p$, we can choose a
basis $(\PCPfrak e^{\ast \alpha}_{(p)},\PCPfrak e^{\ast i}_{(p)})$ of
$T^\ast\PCPcal P\PCPrestrict{\PCPcal U_p}$ as above (in what follows
we will omit the point $p$ when writing a covector). Then we have
\[
   \alpha(p)=\sum_{l=0}^r\alpha_l(p),
\]
where $\alpha_l$ is of the form
\[
   \alpha_l(p)
   =
   \frac1{r!}\alpha_{M_1\cdots M_l i_{l+1}\cdots i_r}(p)\,
   \PCPfrak e^{\ast M_1}\wedge \cdots\wedge \PCPfrak e^{\ast M_l}\wedge 
   \PCPfrak e^{\ast i_{l+1}}\wedge \cdots\wedge \PCPfrak e^{\ast i_r}.
\]
As 
$\PCPfrak e^{\ast M_1}
   \wedge \cdots\wedge \PCPfrak e^{\ast M_l}\wedge 
   \PCPfrak e^{\ast i_{l+1}}
   \wedge \cdots\wedge \PCPfrak e^{\ast i_r}
$
and 
$\PCPfrak e^{\ast M_1}
   \wedge \cdots\wedge \PCPfrak e^{\ast M_j}\wedge 
   \PCPfrak e^{\ast i_{j+1}}\wedge \cdots\wedge \PCPfrak e^{\ast i_r}
$
are linearly independent for $j\neq l$, $d^V\alpha=0$ implies 
\[
   d^V\alpha_l=0 \quad \forall l=1,\ldots,r.
\]
Furthermore, we see that
\[
   d^V\alpha_l(p)
   =0\quad\Leftrightarrow\quad d^V\alpha_{l,i_{l+1}\cdots i_r}(p)=0
   \quad\forall i_{l+1},\cdots, i_r=1,\ldots,n,
\]
where
\[
   \alpha_l(p)=\frac1{(r-l)!}\alpha_{l,i_{l+1}\cdots i_r}(p)\wedge 
   \PCPfrak e^{\ast i_{l+1}}\wedge \cdots\wedge \PCPfrak e^{\ast i_r}.
\]
Now, if we restrict the $\alpha_{l,i_{l+1}\cdots i_r}$ to a fixed
fibre $\PCPcal P_m$ of $\PCPcal P\rightarrow\PCPcal M$, applying $d^V$
corresponds to the exterior derivative on that space. As the fibre
under consideration is a vector space, it follows that 
\[
   \alpha_{l,i_{l+1}\cdots i_r}\PCPrestrict{\PCPcal Z_m}
   =d^V\,\beta^m_{(l-1),i_{l+1}\cdots i_r},
\]
and hence
\[
\begin{split}
  \alpha(p)=\sum_{l=0}^r\alpha_l(p)&=
  \sum_{l=0}^r\frac1{(r-l)!}\alpha_{l,i_{l+1}\cdots i_r}(p)\wedge 
  \PCPfrak e^{\ast i_{l+1}}\wedge \cdots\wedge \PCPfrak e^{\ast i_r}
\\
  &=\sum_{l=0}^r
  \frac1{(r-l)!}
  \left(
    d^V\,\beta^{\pi(p)}_{(l-1),i_{l+1}\cdots i_r}
  \right)\wedge 
  \PCPfrak e^{\ast i_{l+1}}\wedge \cdots\wedge \PCPfrak e^{\ast i_r}
\\
  &=\sum_{l=0}^r
  \frac1{(r-l)!}d^V
  \left(
    \beta^{\pi(p)}_{(l-1),i_{l+1}\cdots i_r}\wedge 
    \PCPfrak e^{\ast i_{l+1}}\wedge \cdots\wedge \PCPfrak e^{\ast i_r}
  \right)
\\
  &=d^V\,\beta(p),
\end{split}
\]
where
\[
   \beta(p)
   =
   \sum_{l=0}^r\frac1{(r-l)!}\beta^{\pi(p)}_{(l-1),i_{l+1}\cdots i_r}\wedge 
   \PCPfrak e^{\ast i_{l+1}}\wedge \cdots\wedge \PCPfrak e^{\ast i_r}.
\]
\PCPqed

\section{Field equations}
As already mentioned the multisymplectic phase space \PCPcal P of a given
field theory is chosen to be the affine dual of the first jet bundle
$\PCPcal J^1\PCPcal V$, but the field equations (\ref{ELG}) are
formulated on $\PCPcal J^1\PCPcal V$ itself. Hence, similar to
ordinary mechanics, one uses the covariant Legendre transformation
(\ref{Leg-Trf}) to reformulate the theory. For this, let us assume
that the middle equation of (\ref{Leg-Koord}) can be rearranged so
that the variables $v^A_i$ can be expressed in terms of
$(x^i,v^A,p^{\PCPstapeln iA})$. In other words, we require 
\begin{equation*}
  \det \left(\frac{\PCPdel^2 L}{\PCPdel v^A_i\PCPdel v^B_j}\right)\neq 0,
  \quad 
  v^A_i=\varphi^A_i(x^i,v^A,p^{\PCPstapeln iA}).
\end{equation*}
Then the Lagrange density $L$, (\ref{Leg-Koord}), becomes a function
over phase space,
\begin{equation*}
  \tilde L(x^i,v^A,p^{\PCPstapeln iA})
  =L(x^i,v^A,\varphi^A_i(x^i,v^A,p^{\PCPstapeln iA}))
\end{equation*}
and we obtain the covariant Hamiltonian
\begin{equation}\label{cov-Ham}
  H(x^i,v^A,p^{\PCPstapeln iA})=
  \tilde L(x^i,v^A,p^{\PCPstapeln iA})-
  p^{\PCPstapeln iA}\varphi^A_i(x^i,v^A,p^{\PCPstapeln iA}).
\end{equation}
Using this, the generalised Hamiltonian equations
\begin{equation}\label{Ham-Gl}
  \frac{\PCPdel H}{\PCPdel v^A}
  =\frac{\PCPdel p^{\PCPstapeln iA}}{\PCPdel x^i},
\quad
  \frac{\PCPdel H}{\PCPdel p^{\PCPstapeln iA}}
  =-\frac{\PCPdel v^A}{\PCPdel x^i},
\end{equation}
are equivalent to the Euler-Lagrange equations (\ref{ELG}), (\cite{Ru66},
ch. 4.2). Note, however, that $H$ is not a function but (\ref{cov-Ham}) rather
describes a subset of \PCPcal P which is the image of $\PCPcal J^1\PCPcal V$
under $\PCPbbm F\PCPcal L$. The coordinates we have used up to now have
arisen in a natural way from coordinates on \PCPcal M and \PCPcal V; they
simply are the components of the tangent map of a given section.
If one uses the connection $\Gamma$ as a zero section of
$\PCPJV\rightarrow\PCPcal V$ one turns \PCPJV into a vector space
$\PCPfrak V\PCPcal V\otimes T^\ast\PCPcal M$, 
and $\PCPcal P$ splits into the direct sum of a line bundle and the
bundle of {\sl linear} mappings of the former vector bundle to
$\Lambda^{n+1}T^\ast\PCPcal M$ (cf. \cite{Sau89}). In coordinates this
corresponds to the change
\begin{equation} 
  \Psi:(x^i,v^A,v^A_i)\mapsto(x^i,v^A,\tilde v^A_i
  =v^A_i+\Gamma_{iB}^A\,v^B).
\end{equation}
Using 
\begin{equation}
  \begin{split}
    \frac{\PCPdel{\PCPcal L}}{\PCPdel v^A_i}\circ\Psi^{-1}
    &=\frac{\PCPdel{\PCPcal L}\circ\Psi^{-1}}{\PCPdel \tilde v^A_i},\\
    \frac{\PCPdel{\PCPcal L}}{\PCPdel v^A}\circ\Psi^{-1}
    &=\frac{\PCPdel\PCPcal L\circ\Psi^{-1}}{\PCPdel \tilde v^A}
    -\Gamma_{iA}^B\frac{\PCPdel\PCPcal L\circ\Psi^{-1}}{\PCPdel {\tilde v}^B_i}
  \end{split}
\end{equation}
equation (\ref{ELG}) becomes (for ${\PCPcal L_\Gamma}=\PCPcal L\circ\Psi^{-1}$)
\begin{equation}
  \frac{\PCPdel\PCPcal L_\Gamma}{\PCPdel v^A}\circ j^1\varphi
  -\nabla_i\left(
    \frac{\PCPdel\PCPcal L_\Gamma}
    {\PCPdel \tilde v^A_i}\circ j^1\varphi
  \right)=0
\end{equation}
For the affine bundle $\PCPcal P$ the change of coordinates induces a
mapping 
$\Psi^\ast:(x^i,v^A,p,p^{\PCPstapeln iA})
  \mapsto
  (x^i,v^A,p+\Gamma_{iB}^A\,p^{\PCPstapeln iA}\,v^B,p^{\PCPstapeln iA})$.
Let $\PCPcal H_\Gamma=H\circ(\Psi^\ast)^{-1}$. As we have a global splitting of
\PCPcal P induced by the connection $\Gamma$, this is a function on
$(\PCPfrak V\PCPcal V\otimes T^\ast\PCPcal M)^\ast$. Differentiating
${\PCPcal H_\Gamma}$ as in (\ref{Ham-Gl}) with respect to $v^A$ and
$p^i_A$ one obtains on solutions $j^1\varphi$ of (\ref{ELG}) 
\begin{equation}
  \frac{\PCPdel \PCPcal H_\Gamma}{\PCPdel v^A}
  =\nabla_i{\tilde p^{\PCPstapeln iA}},
  \quad
  \frac{\PCPdel \PCPcal H_\Gamma}{\PCPdel \tilde p^{\PCPstapeln iA}}
  =-\nabla_i v^A.
\end{equation}
For the last equation we have used that in the coordinates introduced
the first jet prolongation has the form (\ref{specForm}). A similar
result can be found in \cite{GiMaSa99}.
\\ 
Now we are going to formulate the equations of motion in a coordinate
free manner. Let solutions of (\ref{ELG}) be described by
$(n+1)$-vector fields $\PCPxx {n+1}\in\Gamma(\Lambda^{n+1} \PCPTJVast)$ with
$T\bar\pi \PCPxx {n+1}\neq 0$. Further, let 
$\PCPxx {n+1}{}^V=\PCPxx{n+1}-(T\bar\pi \PCPxx {n+1})^h$ 
be the vertical component of $\PCPxx {n+1}$, where 
$(T\bar\pi\PCPxx{n+1})^h$ is the horizontal lift according to the
splitting induced by the mapping (\ref{Zhg-Abb}) in the appendix
\ref{Kov-Abl}. If $\Omega^{(2,n)}=d^V\Theta^{(1,n)}$, where
$\Theta^{(1,n)}$ denotes the vertical component of $\Theta$ (so that
in the splitting above $\Omega^{(2,n)}$ has two vertical and $n$
horizontal components), 
\begin{equation*}
  \Theta^{(1,n)}=\Theta-\Theta^H,
  \quad
  (X)^h\PCPHook\Theta^{(1,n)}=0
  \quad
  \forall\, X\in \Lambda^{n+1} T\PCPcal M,\quad\quad
  X\PCPHook \Theta^H
  =0\quad\forall\quad X\in \PCPfrak V\PCPcal P.
\end{equation*}
the generalised Hamilton equations (\ref{Ham-Gl}) are equivalent to
\begin{equation*}
  \left(X{}^V\PCPHook\Omega^{(2,n)}\right)^{(1,0)}=(-)^{n+1} d^V H.
\end{equation*}

\section{Hamiltonian forms and a graded Poisson structure}

With the help of the vertical exterior derivative we can define the graded
vertical Lie derivative by an $r$-vector field by
\begin{equation}\label{LieDer}
  \PCPcal L_{\PCPxx r}\Phi
  =\PCPxx r\PCPHook d^V\Phi+(-)^{r+1}d^V\left(\PCPxx r\PCPHook\Phi\right)
\end{equation}
for every form $\Phi$ on $T\PCPcal P$.

An $r$-vector field $\PCPxx r$ is called a Hamiltonian multi-vector
field iff there is a horizontal $(n+1-r)$-form $\PCPF {(n+1-r)}$ that
satisfies 
\begin{equation}\label{DefHamVf}
  \PCPxx r\PCPHook\PCPoV=d^V\;\PCPF {(n+1-r)}.
\end{equation}

The set of all such forms will be called the set of Hamiltonian forms and
denoted by $\PCPcal H\PCPcal F$. Not every horizontal form is
automatically Hamiltonian. Indeed, if we write in local coordinates 
\begin{equation}\label{HamFormCoord}
  \PCPF{(n+1-r)}
  =\frac1{r!}F^{i_1\cdots i_r}(\PCPfrak e_{i_1\cdots i_r}\PCPHook\omega),
\end{equation}
where $\omega$ is the horizontally lifted volume form of $\PCPcal M$ and 
$\PCPfrak e_{i_1\cdots i_r}
   =\PCPfrak e_{i_1}\wedge\cdots\wedge\PCPfrak e_{i_r}$, 
we find for $n+1>r$ (\cite{Kan99})
\begin{equation}\label{HamVfCoord}
  \begin{split}
    r\,X^{A[j_1\cdots j_{r-1}}\delta^{i]}_j
    &=\PCPdel_{\PCPsstapeln iA} F^{j_{1}\cdots j_{r-1}i}
\\
    -r\,X^{\PCPistapeln iAj_1\cdots j_{r-1}}
    &=\PCPdel_A F^{j_{1}\cdots j_{r-1}i}
  \end{split}
\end{equation}
which puts a restriction on the admissible horizontal forms $F$ with
$r<n+1$, namely 
\begin{equation}\label{HamFormCond}
  \PCPdel_{\PCPsstapeln {k\phantom{_1}}{B\phantom{_1}}}
  F^{j_{1}\cdots j_{r}}=0
\end{equation}
for all $k\not\in\{j_1,\cdots,j_r\}$. For $r=n+1$ the first equation
in (\ref{HamVfCoord}) does not lead to any restriction, since $j$ has
to be in $\{j_1,\ldots,j_n,i\}$ in any case. Moreover, from 
$d^V \PCPxx r\PCPHook\PCPoV=(d^V)^2\;\PCPF {(n+1-r)}=0$ 
we deduce in particular
\begin{equation*}
  \sum_{i=1}^{n+1}\sum_{A,B=1}^N\PCPdel_{\PCPsstapeln iA}X^{Bi_1\cdots i_r} 
  \PCPfrak e^{\PCPistapeln iA}\wedge\PCPfrak e^{\PCPistapeln jB}
  \wedge \PCPfrak e_{i_1\cdots i_rj}\PCPHook \omega 
  =0,
\end{equation*}
which implies
\begin{equation}
  \left(\PCPdel_{\,\PCPsstapeln {j_1}A}\right)^2
  F^{j_1\cdots j_r}
  =
  -r\PCPdel_{\,\PCPsstapeln {j_1}A}X^{Bj_1\cdots j_{r-1}}=0 
  \textrm{ (No summation over $j_1$.)} 
\end{equation}
Hence, as already remarked in \cite{Kan97}, the coordinate expression
of $F$ can depend on the coordinates of the fibre of \PCPcal P in a
specific polynomial way only, where each coordinate 
$p\,^{\PCPistapeln iA}$ appears at most to the first power.  

If $n=0$ then $\Omega^{(2,0)}$ does not contain any horizontal degree and the
Hamiltonian forms are just functions on \PCPcal P. For those, the conditions
(\ref{HamVfCoord}) become
\begin{equation}\label{n=0-case}
  X^A=\PCPdel_{\PCPistapeln1A} F^1,\quad X^{\PCPistapeln1A}=\PCPdel_A F^1. 
\end{equation}
Hence, arbitrary functions $F$ are allowed. 

\begin{elemma}\label{polyHamForm-lemma}
  Let 
  $\PCPF {(n+1-r)}
     =\frac 1{r!}F^{j_{1}\cdots j_{r}}
     \PCPfrak e_{j_{1}\cdots j_{r}}\PCPHook \omega
  $
  be a Hamiltonian form. If $r<n+1$, then the coefficient functions
  are of the following form. 
  \begin{equation}\label{polyHamForm}
    F^{j_{1}\cdots j_{r}}(x,v,p)
    =\frac 1{r!}\sum_{k=0}^r
    p\,^{\PCPistapeln {j_1}{A_1}}\cdots 
    p\,^{\PCPistapeln {j_k}{A_k}}
    f^{A_1\cdots A_k j_{k+1}\cdots j_r},
  \end{equation}
  where the functions $f$ are antisymmetric in the upper indices.\\
  If $n+1=r$, then the set of Hamiltonian forms consists of all
  functions on the phase space \PCPcal P.
\end{elemma}
With that, we have the following observation.
\begin{elemma}
  If \PCPxx r,\PCPxx s are Hamiltonian multi-vector fields, then 
  \begin{equation}\label{LieBracket}
    \PCPxx r\PCPHook\PCPxx s\PCPHook \PCPoV
  \end{equation}
  is a Hamiltonian form.
\end{elemma}
\textbf{Proof:}
This can be checked by a calculation using coordinates. Let us suppose $n>0$.
(The case $n=0$ is easy because there is no additional restriction on
Hamiltonian forms apart from having horizontal degree zero.) Firstly,
the above expression (\ref{LieBracket}) is horizontal. Since 
$\PCPxx r$ and $\PCPxx s$ are assumed to be Hamiltonian, there are
horizontal forms $F$ and $G$ satisfying (\ref{DefHamVf})
respectively. We will show that 
$\PCPxx r\PCPHook\PCPxx s\PCPHook\PCPoV$ 
is of the form (\ref{polyHamForm}).
\begin{equation*}
  \begin{split}
    \PCPxx r\PCPHook\PCPxx s\PCPHook \PCPoV
    &=
    \frac 1{(r-1)!}\frac 1{(s-1)!}(-)^{(r-1)}
    {\PCPxx r}\rule{0ex}{1.5ex}^{Mi_1\cdots i_{r-1}}
    {\PCPxx s}\rule{0ex}{1.5ex}^{Nj_1\cdots j_{s-1}}
    \langle \PCPfrak e_M\wedge\PCPfrak e_N,
    \PCPfrak e^A\wedge \PCPfrak e^{\PCPistapeln iA}\rangle
    \left(
      \PCPfrak e_{i_1\cdots i_{r-1}j_1\cdots j_{s-1}i}\PCPHook\omega
    \right)
\\
    &=\frac1{(r+s-1)!}H^{i_1\cdots i_{r-1}j_1\cdots j_{s-1}i}
    \left(
      \PCPfrak e_{i_1\cdots i_{r-1}j_1\cdots j_{s-1}i}\PCPHook \omega
    \right).
  \end{split}
\end{equation*}
Because of the special form of $\PCPxx r$ and $\PCPxx s$ according to lemma
\ref{polyHamForm-lemma} we find
\begin{equation}
  \PCPdel_{\,\PCPsstapeln {i_1}A}H^{i_1\cdots i_{r+s-1}}
  =-\PCPdel_{\,\PCPsstapeln {i_2}A}H^{i_1\cdots i_{r+s-1}}
\end{equation}
and
\begin{equation}
  \PCPdel_{\PCPsstapeln {i}A}H^{i_1\cdots i_{r+s-1}}
  =0 \quad 
  \textrm{ for } i\not\in\{i_1,\cdots, i_{r+s-1}\}.
\end{equation}
This shows that $\PCPxx r\PCPHook\PCPxx s\PCPHook\PCPoV$ fulfils the
conditions derived from (\ref{HamVfCoord}) and thus is Hamiltonian. 
\PCPqed
Looking at equation (\ref{LieBracket}) we can ask what the corresponding
Hamiltonian multi-vector field might be. One calculates
\begin{equation*}
  \begin{split}
    d^V\left(\PCPxx r\PCPHook\PCPxx s\PCPHook\PCPoV\right)&=
    d^V\left(\PCPxx r\PCPHook\PCPxx s\PCPHook \PCPoV\right)+
    (-)^{r+1}
    \PCPxx r\PCPHook d^V\left(\PCPxx s\PCPHook \PCPoV\right)
\\
    &=\PCPcal L_{\PCPxx r}\PCPxx s\PCPHook\PCPoV
  \end{split}
\end{equation*}
As $\PCPcal L_{\PCPxx r}\PCPoV=0$ this looks like the Lie bracket of
$\PCPxx r$ and $\PCPxx s$ being inserted in \PCPoV. Now in symplectic
mechanics the Lie bracket of two (locally) Hamiltonian vector fields
is the vector field associated to the Poisson bracket of the
Hamiltonian functions of the former. Hence, by analogy, we define a
bracket as follows: 
\begin{equation}\label{PoissBra}
  \{\PCPF r,\PCPF s\}
  =(-)^{n+1-r}\PCPxx {n+1-r}\PCPHook\PCPxx{n+1-s}\PCPHook\PCPoV,
\end{equation}
where $\PCPF r,\PCPF s$ are Hamiltonian forms and 
$\PCPxx{n-r},\PCPxx{n-s}$ denote the corresponding vector fields. Note
that whereas there is some ambiguity in the choice of a Hamiltonian
(multi-)vector field in eq. (\ref{DefHamVf}), this does not lead to an
ambiguity of the above bracket. Indeed, a vector field $X$ that
vanishes on $\PCPoV$ must have vanishing coefficients 
$X^{M i_1 \cdots i_k}$ but can have non vanishing components 
$X^{M_1\cdots M_j i_1\cdots i_l}$. The latter, however, do not
contribute to the bracket since $\PCPoV$ is of type $(2,n)$\footnote{%
  The author wishes to thank the referees for pointing out the
  remaining ambiguity to him.
}.%
\begin{prop}\label{PoissStr}
  The bracket 
  \begin{equation}
    \{\cdot,\cdot\}:
    \PCPcal H\PCPcal F\times\PCPcal H\PCPcal F
    \rightarrow
    \PCPcal H\PCPcal F
  \end{equation}
  defined by (\ref{PoissBra}) has the following properties:
  \begin{enumerate}
  \item It is graded antisymmetric,
    \begin{equation*}
      \{\PCPF r,\PCPF s\}=-(-)^{(n-r)(n-s)}\{\PCPF s,\PCPF r\}.
    \end{equation*}
  \item
    It fulfils a graded Jacobi identity,
    \begin{equation*}
      (-)^{(n-r)(n-t)}\{\PCPF r, \{\PCPF s,\PCPF t\}\}
      +(-)^{(n-s)(n-r)}\{\PCPF s, \{\PCPF t,\PCPF r\}\}
      +(-)^{(n-t)(n-s)}\{\PCPF t, \{\PCPF r,\PCPF s\}\}=0.
    \end{equation*}
  \item
    There is a product
    \begin{equation}\label{superprod}
      \PCPF r\bullet\PCPF s
      =\ast^{-1}\left(\ast\PCPF r\wedge\ast\PCPF s\right)
      =(-)^{(n+1-r)(n+1-s)}\PCPF s\bullet\PCPF r,
    \end{equation}
    where $\ast$ is the operation induced by the Hodge operator on \PCPcal M 
    that maps Hamiltonian functions to Hamiltonian functions. With respect to
    $\bullet$, the above defined bracket shows a graded Leibniz rule,
    \begin{equation}\label{Leibniz}
      \{\PCPF r,\PCPF s\bullet \PCPF t\}
      =\{\PCPF r,\PCPF s\}\bullet \PCPF t+
      (-)^{(n-r)(n+1-s)}\PCPF s\bullet \{\PCPF r,\PCPF t\}.
    \end{equation}
  \end{enumerate}
\end{prop}
{\bf Proof.} ${\it 1)}$ is an immediate consequence of the definition.\\
${\it 2)}$ is a straightforward calculation if one uses 
\begin{equation}\label{folg-HamFormCond}
  \PCPdel_{\PCPsstapeln kB}X^{\PCPistapeln iAj_1\cdots j_{-1}}
  =-\PCPdel_A X^{b[j_1\cdots j_{-1}}\delta^{i]}_k,
  \quad
  \PCPdel_BX^{\PCPistapeln iAj_1\cdots j_{-1}}
  =\PCPdel_AX^{\PCPistapeln iBj_1\cdots j_{-1}}
\end{equation}
which can be deduced from changing the order of differentiation in
(\ref{HamVfCoord}).\\
As for ${\it 3)}$, using 
\begin{equation*}
  \ast\left(
    \PCPfrak e_{i_1\cdots i_r}\PCPHook 
    \PCPfrak e^1\wedge\cdots\wedge \PCPfrak e^n
  \right)
  =
  \PCPfrak e^{i_1}\wedge\cdots\wedge \PCPfrak e^{i_r}
\end{equation*}
we find 
\begin{equation}\label{bullet-local}
  \PCPG {n+1-q}\bullet \PCPHH {n+1-r}
  =\frac 1{(q+r)!}G^{i_1\cdots i_q}H^{i_{q+1}\cdots i_{q+r}}
  \left(\PCPfrak e_{{i_1}\cdots i_{q+r}}\PCPHook\omega\right)
\end{equation}
and hence
\begin{equation}
  \begin{split}
    \{\PCPF {n+1-p},\PCPG {n+1-q}\bullet\PCPHH {n+1-r}\}
    &=(-)^p
    \frac 1{(p-1)!} 
    X_F^{Mi_1\cdots i_{p-1}}\PCPHook 
    d^V\left( G\bullet H\right)
    \\
    &=
    X_F^{Mi_1\cdots i_{p-1}}
    (\PCPdel_M G^{j_1\cdots j_{q}}H^{j_{q+1}\cdots j_{q+r}}
    \PCPfrak e_{i_1\cdots i_{(p-1)}j_1\cdots  j_{q+r}}\PCPHook\omega
    \\
    &\quad+
    (-)^{(p-1)q}
    G^{j_1\cdots j_q}
    X_F^{Mi_1\cdots i_{p-1}}(\PCPdel_M H^{j_{q+1}\cdots j_{q+r}})
    \PCPfrak e_{j_1\cdots j_qi_1\cdots i_{(p-1)}j_1\cdots  j_{q+r}}
    \PCPHook\omega
    \\
    &=\{\PCPF p,\PCPG q\}\bullet\PCPHH r
    +(-)^{(p-1)q}\PCPG q\bullet\{\PCPF p,\PCPHH r\}
  \end{split}
\end{equation}
\PCPqed
One might ask about the dependence of the bracket on the connections
$\Gamma$ and $\Lambda$. As can be seen from (\ref{split}), different
choices of connections amount to differences in the horizontal terms
of the vertical forms that have been used in the definition of
$d^V$. But from (\ref{DefHamVf}) we learn that this change can have an
effect on those terms of $X$ that have two or more vertical components
only. Again, those terms do not contribute to the bracket. Hence the
Poisson bracket does not depend on $\Gamma$ nor $\Lambda$.

\section{Recovering mechanics}

To recover Hamiltonian mechanics we proceed as follows. Let \PCPcal Q be the
coordinate space of the theory. Then, $\PCPcal M=\PCPbbm R$ and 
\PCPcal V is trivial $\PCPcal V=\PCPbbm R\times\PCPcal Q$. Hence, 
$T\PCPcal V$ decomposes into
$T\PCPcal V=\PCPbbm R\oplus T\PCPcal Q$. The condition for a mapping 
$\varphi\oplus\psi:T\PCPcal M
   =\PCPbbm R\rightarrow T\PCPcal V=\PCPbbm R\oplus T\PCPcal Q
$ 
to be in \PCPJV is thus
\begin{equation}
  T\pi\circ(\varphi\oplus\psi)=\psi=T\textrm{id}_{\PCPbbm R}=1.
\end{equation}
As the mapping $\varphi$ is defined by its value at $1$ we conclude
$\PCPJV=T\PCPcal Q\times \PCPbbm R$ and, going to the dual we obtain
the phase space, 
\begin{equation}
  \PCPcal P\PCPJVast=(T^\ast\PCPcal Q\oplus\PCPbbm R)\times\PCPbbm R.
\end{equation}
The canonical $1$-form $\Theta$ reads
\begin{equation*}
  \Theta(t,v^A,p,p_A)=p_A\,dv^A+p\,dt
\end{equation*}
whereas $\Omega^{(2,0)}$ is
\begin{equation*}
  \Omega^{(2,0)}(t,v^A,p,p_A)=dp_A\wedge dv^A
\end{equation*}
which is just the canonical $2$-Form. As the base manifold is one-dimensional,
horizontal forms are either functions or 1-forms on $T^\ast\PCPcal Q$.
Now in this case equation (\ref{DefHamVf}) admits the former case since
$\Omega^{(2,0)}$ does not contain any horizontal component. Therefore
the Hamiltonian multi-vector fields can be ordinary vector fields on
$T^\ast\PCPcal Q$ only, and we have
\begin{equation}
  X_F(t,v,p)
  =\PCPdel_{p^A}F(t,v,p)\PCPdel_{p^A}
  -\PCPdel_{v^A}F(t,v,p)\PCPdel_{v^A}
\end{equation}
There is no additional restriction to admissible Hamiltonian functions
(cf. (\ref{n=0-case})) and we have arrived at the stage of Hamiltonian
mechanics (cf. \cite{Gue91}). As the bundle \PCPcal V is trivial we do
not need a connection really, so there is no need for \PCPcal Q to be
a vector bundle. As the base manifold is one-dimensional only, the
product of two Hamiltonian forms always gives zero. This can be
remedied if one includes  horizontal $1$-forms in the set of
observables in addition to functions\footnote{%
I. Kanatchikov, private communication.}.
This leads to the extension of the notion of Hamiltonian vector fields to form
valued vector fields.
\\ 
In \cite{GoSt73}, sec. 4, where a Poisson structure is defined on (de
Rham) equivalence classes of forms on \PCPcal P, the Poisson algebra
consists of those functions only for which the dependence on the
parameter is the physical time, i.e. which solve the equations of
motion when differentiated with respect to this parameter. Here, in
contrast, nothing can be said about the "time" dependence of
Hamiltonian forms.

\section{The case of a scalar field}

In the case of a scalar field, the fibre of \PCPcal V is isomorphic to
\PCPbbm R. Using a connection $\Gamma:\PCPcal V\rightarrow\PCPJV$, we
obtain an isomorphism 
\begin{equation}
  \PCPJV\stackrel\Gamma
  \cong \PCPfrak V\PCPcal V\otimes_{\PCPcal V} T^\ast\PCPcal M,
  \quad \PCPfrak V\PCPcal V\cong\PCPbbm R\times\PCPbbm R.
\end{equation}
Hence
\begin{equation}
  \PCPJV\stackrel\Gamma\cong \textrm{pr}^\ast (T^\ast \PCPcal M),
\end{equation}
where pr denotes the canonical projection of the bundle 
$\PCPcal V\rightarrow\PCPcal M$. Using (\ref{DefHamVf}) one
immediately verifies in local coordinates $(x^i,v,p^i,p)$ of \PCPcal P
in this case (let $e_i$ denote the horizontal lifts of tangent vectors
of \PCPcal M and $\PCPfrak e^i$ be the vertical forms with respect to
the splitting discussed in the appendix; the determinant comes from
the volume element on \PCPcal M) 
\begin{center}
    $-\PCPdel_v\PCPHook\Omega^{(2,n)}
    =\PCPfrak e^i\wedge(e_{i}\PCPHook\omega)
    =d^V p^i\wedge(e_{i}\PCPHook\omega),$
\\
    $\sum_{i=1}^{n+1}\PCPdel_{p^i}\wedge({(-)^i(\sqrt{\det g})}
    e_{1}\wedge\cdots\wedge
    e_{{i-1}}\wedge e_{{i+1}}\wedge\cdots\wedge e_{{n+1}})\PCPHook
    \Omega^{(2,n)}
    =d^Vv,$
\end{center}
hence $\Pi(x,v,p)=p^i\wedge(e_{i}\PCPHook\omega)$ satisfies 
\begin{equation*}
  \{\Pi,\Phi\}=1
\end{equation*}
for $\Phi(x,v)=v$, but $\Pi\bullet 1 =0=\Phi\bullet 1$. The unit with
respect to $\bullet$ is $\omega$, so one should look for solutions of
\begin{eqnarray*}
  X\PCPHook Y\PCPHook\PCPoV = \omega.
\end{eqnarray*}
This cannot be solved, as $\PCPoV$ contains $n$ horizontal components,
whereas $\omega$ is a horizontal $(n+1)$-form. As before, one might
have to include vector fields that are form valued, i.e. endomorphisms
of $\Lambda^\ast T^\ast\PCPcal P$.
\\ 
Note, however, that the connection $\Gamma$ remains arbitrary: 
Although it is hidden in the expression for $\Pi$,
\begin{equation*}
  \Pi(x,v,p)
  =p^i\wedge(e_i\PCPHook\omega)=p^i\wedge(\PCPdel_i\PCPHook\omega),
\end{equation*}
$\Pi$ is in fact independent of it.

\section{Conclusions}

In multisymplectic geometry we take the phase space \PCPcal P to be
the affine dual of the first jet bundle to a given vector bundle
\PCPcal V. It is then possible to define (graded) Poisson brackets
(\ref{PoissBra}) on \PCPcal P even for nontrivial vector bundles. For
this one needs a covariant derivative on the $(n+1)$-dimensional base
manifold \PCPcal M (space-time) and a connection on the vector bundle
of the fields under consideration.

Kanatchikov has proposed a similar construction by making use of 
equivalence classes of forms modulo forms of higher horizontal degree
(\cite{Kan98}). This is equivalent to the use of the construction
elaborated in this article, as a vertical form, say $\PCPfrak e^A$,
differs from the coordinate expression $dv^A$ by horizontal components
only, cf. (\ref{split}),
\begin{equation}
  \PCPfrak e^{\ast A}_{(p)}=dv^A+\Gamma_{iB}^A(\pi(p))\,v^B\,d x^i.
\end{equation}
Hence, $\PCPfrak e^{\ast A}_{(p)}$ and $dv^A$ define the same
equivalence class, independent of the connections $\Gamma$ and
$\Lambda$ used. The same applies to the bracket: Whereas the
correspondence of Hamiltonian forms and multi-vector fields is
ambiguous and does depend on the connections chosen, the (graded)
Poisson bracket does not. Admissible observables are so-called
Hamiltonian forms, horizontal forms that satisfy certain consistency
relations, (\ref{HamVfCoord}). It turns out that those Hamiltonian
forms are polynomial in the momenta, i.e. coordinates of the fibres of
$\PCPcal P\rightarrow\PCPcal V$, cf. (\ref{polyHamForm}). 

In addition \PCPcal M has to be orientable in order to define the
multiplication (\ref{superprod}) between Hamiltonian forms. For
Hamiltonian forms of the same degree, this product is commutative but
gives zero if the form degree is less than $(n+1)/2$.

If space-time is taken to be one-dimensional the whole formalism
reduces to ordinary mechanics on a configuration space \PCPcal Q. 
Hamiltonian forms then are arbitrary functions on the extended phase 
space $T^\ast\PCPcal Q\times \PCPbbm R$, and the Poisson bracket takes the
standard form. However, the product $\bullet$ of functions always
gives zero in this case.

In the case of a scalar field, given a (local) field $\Phi$ one can
define a Hamiltonian form $\Pi$ that satisfies $\{\Pi,\Phi\}=1$, but
the constant function $1$ is not the unit with respect to
$\bullet$. Rather, this r\^ole is played by $\omega$, the pulled back
volume form from \PCPcal M. To obtain $\{\Pi,\Phi\}=\omega$ one has to
extend the notion of Hamiltonian vector fields in a way similar to
that needed in the mechanical case (as mentioned above), namely one
has to include form valued vector fields, i.e. endomorphisms of
$\Lambda^\ast T^\ast \PCPcal P$. 

The Poisson structure is graded in the following way. Let the degree
of a (homogeneous) Hamiltonian form be its degree as an element of the
exterior algebra. Then the degree of the Poisson bracket of two
Hamiltonian forms is the sum of the respective degree minus $n$, the
number of space directions, while the degree of the product of two
Hamiltonian forms is the sum of the degrees minus $n+1$,
\begin{equation}
  \deg \{\PCPF r,\PCPF s\}=\deg \PCPF r+\deg\PCPF s-n,\quad
  \deg \PCPF r\bullet \PCPF s=\deg \PCPF r+\deg\PCPF s-(n+1).
\end{equation}
Looking at proposition \ref{PoissStr} we find that the graded
antisymmetry of the $\{,\}$, the graded Jacobi identity, the graded derivation
property with respect to $\bullet$ and the graded commutativity of
$\bullet$ all match with each other.

As already remarked in the examples, How to relate observables of
physical fields and Hamiltonian forms. This point requires further
investigation, especially the relation with the multiplicative
structure. In particular, the notion of canonical conjugate momenta
needs to be clarified. 
\\[2ex]
{\bf Note added.} 
As pointed out by one of the referees the above construction depends
heavily on the vector space structure of fibres of \PCPcal V. This
might be sufficient for the study of such field theories where the
fields take their values in a vector space. For classical mechanics on
arbitrary configurations spaces, nevertheless, or in the case of
string theory -- whenever the target space is not Minkowski space --
there's is indeed a need for a generalisation of the construction. In
this article, all that is used really is a splitting of the tangent
space $T\PCPfrak J^1\PCPcal V$ in horizontal and vertical subspaces
with respect to the canonical projection onto \PCPcal M.
Such a splitting does not exist canonically. There is, however, a
natural way to split 
${(\pi_1)^1_0}^\ast \left(T\PCPfrak J^1\PCPcal V\right)$,  
the pull back of $T\PCPfrak J^1\PCPcal V$ onto 
$\PCPfrak J^1\PCPfrak J^1\PCPcal V$, the first jet bundle of 
$\PCPfrak J^1\PCPcal V$. Now every connection 
$\bar\Gamma$ on $\PCPfrak J^1\PCPcal V$ (viewed as a bundle over \PCPcal M) 
defines a map 
$\bar\Gamma:\PCPfrak J^1\PCPcal V
   \rightarrow\PCPfrak J^1\PCPfrak J^1\PCPcal V$ 
and hence induces a splitting of $T\PCPfrak J^1\PCPcal V$. For \PCPcal
V being a general fibre bundle, the connection $\Gamma$ does not 
depend linearly on the fibre coordinates (cf. (\ref{conn})). Rather,
it takes the most general form
\begin{equation*}
  \Gamma:\PCPcal V\ni(x^i,u^A)\mapsto (x^i,u^A,\Gamma_i^A).
\end{equation*}
In this case, in the local expression (\ref{eq:Gamma-bar}), one has to replace
$-\Gamma_{iB}^Au^B_j$ by $\PCPdel_{u^B}\Gamma_i^A u^B_j$ and 
$\Gamma^A_{kB}u^B$ by $-\Gamma_k^A$.
\\[1ex]
{\bf Acknowledgements.} The author's interest in this subject was initiated by
very elucidating discussions with H. R\"omer and M. Bordemann about
quantisation schemes for field theories. In particular, the author thanks
M. Bordemann for explaining \cite{BNW98a} to him and for critical
remarks. Finally clarifying discussions with and valuable comments by
I. Kanatchikov are gratefully acknowledged.

\begin{appendix}
\section{Connections and jet bundles\label{Zhg+Jetb}}

Given a bundle $\pi:\PCPcal V\rightarrow\PCPcal M$ over an $n$-dimensional base
manifold \PCPcal M every connection is defined by a section $\Gamma$
of the first jet bundle $\PCPfrak J^1\PCPcal V$ of $\PCPcal V$, since
it describes how to lift tangent vectors of the base manifold
horizontally. If in addition \PCPcal V is a vector bundle (with fibre
$V$) then as $\PCPfrak J^1\PCPcal V$ is an affine bundle over 
\PCPcal V the connection $\Gamma$ delivers an isomorphism
\begin{equation}\label{jet-as-vb}
  \PCPfrak J^1\PCPcal V
  \stackrel{\Gamma}{\cong}\PCPfrak V\PCPcal V\otimes_{\PCPcal M}
  T^\ast\PCPcal M,
\end{equation}
where both sides ($\PCPfrak V\PCPcal V$ being the vertical bundle to
\PCPcal V) are viewed as bundles over the base manifold \PCPcal
M. Note in particular that the vertical bundle $\PCPfrak V\PCPcal V$
is a vector bundle over \PCPcal M (with typical fibre $V\times V$,
\cite{KMS93}, ch. II, 6.11.).
\\
Now for \PCPcal V being a vector bundle we can form the covariant derivative
$\nabla$  that corresponds to the given connection $\Gamma$. Then horizontal
lifts of tangent vectors are represented by covariantly constant lifts
of curves in the base manifold \PCPcal M. Therefore, in local coordinates
$(x^i)_{i=1,\ldots,n}$ of  \PCPcal M and 
$(x^i,v^A)_{i=1,\ldots,n,A=1,\ldots,N}$ of
\PCPcal V the map 
$\Gamma(v)\in\left(\PCPfrak J^1\PCPcal V\right)_v$, $v\in\PCPcal V$, takes 
the form
\begin{equation}\label{conn}
  \Gamma(v):\; (x,\dot c^i(x))
  \mapsto
  \left(x,v,-\Gamma_{iB}^A(x)\,v^B\right),
\end{equation}
where $\Gamma_{iB}^A(x)$ is the Christoffel symbol of $\nabla$.

Now we are locking for a connection in $\PCPfrak J^1\PCPcal V$, that
is for a map 
\begin{equation*}
  \bar\Gamma:\;\PCPfrak J^1\PCPcal V
  \rightarrow\PCPfrak J^1\left(\PCPfrak J^1\PCPcal V\right).
\end{equation*}
For this, one needs a connection both in \PCPcal V and \PCPcal M (\cite{Kol87},
Prop. 4). If we use the isomorphisms
\begin{equation*}
  \PCPfrak J^1\PCPcal V
  \stackrel\Gamma\cong \PCPfrak V\PCPcal V\otimes T^\ast\PCPcal M
  \quad
  \textrm{and}
  \quad
  \PCPfrak J^1\left(\PCPfrak V\PCPcal V\otimes T^\ast\PCPcal M\right)
  \cong
  \PCPfrak J^1
  \left(
    \PCPfrak V\PCPcal V
  \right)
  \otimes\PCPfrak J^1\left(T^\ast\PCPcal M\right),
\end{equation*}
the latter being natural, we see that all we need is a map 
$\PCPfrak V\PCPcal V\rightarrow\PCPfrak J^1\PCPfrak V\PCPcal V$, 
since a connection on \PCPcal M defines a map 
$\Lambda^\ast:T^\ast\PCPcal M
   \rightarrow\PCPfrak J^1\left(T^\ast\PCPcal M\right)$. Now
the desired map can be constructed by vertical prolongation if we make use of
the isomorphism 
$\PCPfrak V\PCPJV\cong\PCPfrak J^1\PCPfrak V\PCPcal V$ (\cite{GoSt73},
eq. (1.4))\footnote{%
  Let $s_t$ denote a one-parameter family of local sections of $\pi:\PCPcal
  V\rightarrow\PCPcal M$. Then 
  \[
    \PCPdbyd t j^1(s_t)(x)\mapsto j^1(\PCPdbyd t s_t)(x)
  \]
  gives the isomorphism.
}:%
\begin{equation*}
  \PCPfrak V\Gamma:\PCPfrak V\PCPcal V\rightarrow
  \PCPfrak V\PCPfrak J^1\PCPcal V\cong\PCPfrak J^1\PCPfrak V\PCPcal V.
\end{equation*}
Indeed,
\begin{equation*}
  \PCPfrak V\Gamma\otimes\Lambda^\ast:
  \PCPfrak V\PCPcal V\otimes T^\ast\PCPcal M
  \rightarrow
  \PCPfrak J^1\PCPfrak V\PCPcal V\otimes\PCPfrak J^1 T^\ast\PCPcal M
  \cong
  \PCPfrak J^1(\PCPfrak V\PCPcal V\otimes T^\ast\PCPcal M)
\end{equation*}
gives a connection\footnote{
In \cite{Kol87}, p. 136, this construction is denoted by $p(\Gamma,\Lambda)$.
}
\begin{equation}\label{Jet-Zhg}
  \bar\Gamma:\PCPfrak J^1\PCPcal V
  \rightarrow\PCPfrak J^1\left(\PCPfrak J^1\PCPcal V\right).
\end{equation}
In coordinates $(x^i,v^A,v^A_i)$ of $\PCPfrak J^1\PCPcal V$ one calculates
\begin{equation}\label{Jet-Zhg-Koord}
  \bar\Gamma(x^i,v^A,v^A_i):
  \left(x^i,\dot x^i\right)
  \mapsto
  \left(
    x^i,v^A,v^A_i,\dot x^i,-\Gamma_{jB}^A(x)\,v^B\,\dot x^j,
    \bar\Gamma_{ij}^A\dot x^j
  \right),
\end{equation}
where
\begin{equation}\label{eq:Gamma-bar}
  \bar\Gamma_{ij}^A(x^i,u^A,u^A_i)=
  -\Gamma_{jB}^A(u^B_i+\Gamma_{iC}^Bu^C)
  -\Lambda_{ji}^k(u^A_k+\Gamma_{kB}^Au^B)
  -(\PCPdel_j\Gamma^A_{iB})u^B+\Gamma_{iB}^A\Gamma_{jC}^Bu^C.
\end{equation}
Note that $\Lambda_{ij}^k$ denote the Christoffel symbols of
$\Lambda$, not $\Lambda^\ast$.

\section{A covariant derivative on $T\PCPcal P$\label{Kov-Abl}}

Using a connection $\Gamma$ of $\pi:\PCPcal V\rightarrow\PCPcal M$,
which is a map 
\begin{equation*}
  \Gamma:\PCPcal V\rightarrow\PCPJV,
\end{equation*}
the affine bundle $\pi':\PCPJV\rightarrow\PCPcal V$ becomes a vector bundle,
\begin{equation*}
  \PCPJV\stackrel\Gamma\cong\PCPfrak V\PCPcal V\otimes_{\PCPcal V}
  \pi^\ast\left(T^\ast\PCPcal M\right),
\end{equation*}
where $\Gamma(\PCPcal V)$ is identified with the zero section.

If in addition $\pi$ is a vector bundle, then \PCPVV is a vector
bundle over \PCPcal M as well (\cite{KMS93}, ch. II, 6.11), and we have
\begin{equation*}
  \PCPJV\stackrel\Gamma\cong\PCPVV\otimes_{\PCPcal M} T^\ast\PCPcal M.
\end{equation*}
Let $\PCPbV=\PCPVV\otimes_{\PCPcal M} T^\ast\PCPcal M$. In
multisymplectic geometry the phase space \PCPJVast consists of all
with respect to $\pi'$ fibre-wise affine mappings from \PCPJV to
$\Lambda^n T^\ast\PCPcal M$. In order to simplify the notation, let us
denote this bundle by $\PCPcal P:=\PCPJVast$. 
Again, the connection $\Gamma$ provides an isomorphism
\begin{equation*}
  \PCPcal P\stackrel\Gamma
  \cong(\PCPbVast\otimes\Lambda^nT^\ast\PCPcal M)
  \oplus_{\PCPcal V}\PCPbbm R,
\end{equation*}
where $\tilde p\in\PCPcal P$ is decomposed into a linear map 
$\bar p:\PCPbV\rightarrow\Lambda^nT^\ast\PCPcal M$ 
and a function $p$ on \PCPcal V in the following way:
\begin{equation*}
  \begin{split}
    \tilde p(\tilde v)&=\tilde p(\tilde v)-\tilde p(\Gamma(\pi'(\tilde v)))+
    \tilde p(\Gamma(\pi'(\tilde v)))
\\
    &=\bar p(\PCPbv)+p(v).
  \end{split}
\end{equation*}
Making use of the duality of \PCPbVast and \PCPbV, we obtain a
connection \PCPbGast on \PCPbVast by 
\begin{equation*}
  \langle \PCPbGast(v),\PCPbv\rangle=\langle\bar p,\PCPbG(v)\rangle,
  \quad
  \forall v\in\PCPcal V,\PCPbv \in \PCPbV_v,\bar p\in\PCPbVast_v.
\end{equation*}
Here, $\PCPbG$ is the connection on \PCPbV as explained in detail in
(\ref{Zhg+Jetb}). Further, this gives a connection on $\PCPcal P$. In
coordinates $(x^i,v^A,p^i_A,p)$ we calculate 
\begin{equation*}
  \PCPbGast(\bar p):T_x\PCPcal M\ni(x^i,\xi^i)
  \mapsto
  (x^i,-\Gamma_{iB}^A \,v^B\,\xi^i,
  (\Lambda_{ji}^kp^i_A-\Gamma_{jA}^Bp^k_B)\xi^j,0)\in T\PCPcal P.
\end{equation*}
Now \PCPbGast defines a covariant derivative $\bar\nabla$ on \PCPcal
P. With the help of this we define the connection mapping $K$ for 
$[\alpha]_p]\in T_p\PCPcal P$, represented by a curve $\alpha(t)$, by
\begin{equation}\label{Zhg-Abb}
  K:T_p\PCPcal P\ni[\alpha]_p\mapsto\left\{
    \begin{array}{ll}
      \PCPdbyd t\alpha(t) &\textrm{ if } T\bar\pi[\alpha]=0\\[1.5ex]
      \left(\bar\nabla_{T\bar\pi[\alpha]}\alpha\right)(0)
      & \textrm{ otherwise. }
    \end{array}\right.
\end{equation}
One easily verifies that $K$ is well defined. Let $p$ be a point in
\PCPcal P and $x$ its image under the projection $\bar\pi$. 
For the tangent mapping of the canonical projection 
$\bar\pi:\PCPcal P\rightarrow\PCPcal M$, the map $K\oplus
T\bar\pi:T_p\PCPcal P\rightarrow\PCPcal P_{x}\oplus T_{x}\PCPcal M$ 
is bijective and hence provides a splitting of $T\PCPcal P_p$. 
$X^h_p\in T_p\PCPcal P$
is called the horizontal lift of $H\in T_x\PCPcal M$ iff 
$K\oplus T\bar\pi(X^h_p)=X$. 
Similarly, $q^v_p\in T_p\PCPcal P$ is called the vertical lift of
$q\in \PCPcal P_x$ iff $K\oplus T\bar\pi(q^v_p)=q$.
Using this we define a covariant derivative $D$ on $T\PCPcal P$ 
by\footnote{%
  This
  method is inspired by the construction in \cite{BNW98a}.
  }:
\begin{equation}
  \begin{split}
    D_{X^h}Y^h\PCPatpoint p
    &=\left(\nabla^{\PCPcal M}_XY\right)^h\PCPatpoint p
    +\frac12\left(\bar R(X,Y)p\right)^v\PCPatpoint p\\
    D_{X^h}\beta^v\PCPatpoint p
    &=\left(\bar\nabla_X\beta\right)^v\PCPatpoint p\\
    D_{\beta^v}X^h\PCPatpoint p
    &=0=
    D_{\beta^v}\Gamma^v\PCPatpoint p,
  \end{split}
\end{equation}
where $p\in\PCPcal P$, 
$\beta^v,\Gamma^v,X^h,Y^h\in T\PCPcal P$ are lifts as above, and 
$\nabla^{\PCPcal M}$ is the (torsion free) covariant derivative on
$T\PCPcal M$. The curvature term $\bar R$ of $\bar\nabla$ is needed
for $D$ to be torsion free. 

Since at every point $p$ of \PCPcal P the tangent space $T_p\PCPcal P$
decomposes into the direct sum of horizontal and vertical vectors, we
can choose an appropriate basis as follows.
If $(x^i)$ are coordinates of a neighbourhood \PCPcal U of \PCPcal M
that trivialises $\PCPcal P\PCPrestrict{\PCPcal U}$ and 
$(\xi^i,v^A,p^{\PCPstapeln iA},p)$ are coordinates on \PCPcal P, we
define for every $p\in\PCPcal P$  
\begin{equation*}
  \begin{split} 
    \PCPfrak e_i(p)
    &=\left(\PCPdel_{x^i}\right)^h\PCPatpoint p=\PCPdel_{\xi^i}
    -\Gamma_{iA}^B\,v^A\,\PCPdel_{v^B}
    +\bar\Gamma_{ij}^A\,\PCPdel_{p^{\PCPstapeln jA}} 
    \\
    \PCPfrak e_A(p)
    &=\PCPdel_{v^A},\quad\PCPfrak e_{\PCPsstapeln iA}(p)
    =\PCPdel_{p^{\PCPstapeln iA}},
    \quad \PCPfrak e(p)=\PCPdel_p,\quad i=1,\ldots,n, \,A=1,\ldots,N.
  \end{split}
\end{equation*}
we obtain a basis of $T_p\PCPcal P$. From the definition of $D$ it
follows in particular that
\[
   D_{\PCPfrak e_A}\PCPfrak e_\alpha=0,
   \quad D_{\PCPfrak e_{\PCPsstapeln iA}}\PCPfrak e_\alpha=0,
   \quad \forall
   \alpha=i,A,\PCPstapeln jB,\quad A,B=1,\ldots,N,\quad i,j=1,\ldots,n.
\]

\end{appendix}


\begin{thebibliography}{10}

\bibitem {BFFLS78}
  {\sc Bayen, F., Flato, M., Fr{{\o}}nsdal, C., Lichnerowicz, A., Sternheimer,
    D.: }\newblock {\em Deformation Theory and Quantization}.
  \newblock Ann. Phys.  {\bf 111} (1978), 61--151.

\bibitem {BNW98a}
  {\sc Bordemann, M., Neumaier, N., Waldmann, S.: }\newblock {\em Homogeneous
    Fedosov Star Products on Cotangent Bundles I: Weyl and Standard
    Ordering with Differential Operator Representation}.
\newblock Commun. Math. Phys.  {\bf 198} (1998), 363--396.

\bibitem {Di91}
  {\sc Dickey, L. A.:}\newblock{\em Soliton Equations and Hamiltonian
    Systems.}
  \newblock Vol. 12 of {\em Advanced Series in Mathematical Physics},
  World Scientific, Singapore, 1991.
  
\bibitem {Dit90}
  {\sc Dito, J.: }\newblock {\em Star-Product Approach to Quantum Field Theory:
    The Free Scalar Field}.
  \newblock Lett. Math. Phys.  {\bf 20} (1990), 125--134.
  
\bibitem {GiMaSa99}
  {\sc Giachetta, G., Mangiarotti, L., Sardanashvily, G.: }\newblock {\em
    Covariant {H}amiltonian equations for field theory}.
  \newblock J. Phys. A {\bf 1(4)} (1999), 375--390.
  \newblock A similar version can be found at
  {\tt hep-th/9904062} under the title "Covariant Hamiltonian Field Theory".

\bibitem {GoSt73}
  {\sc Goldschmidt, H., Sternberg, S.: }\newblock {\em The {H}amilton-{C}artan
    formalism in the calculus of variations}.
  \newblock Ann. Inst. Fourier  {\bf 23}.1 (1973), 203--267.
  
\bibitem {GoIsMa98}
  {\sc Gotay, M.~J., Isenberg, J., Marsden, J.~E.: }\newblock 
  {\em
    Momentum Maps and Classical Relativistic Fields I: Covariant Field
    Theory
    },
  \newblock {\tt physics/9801019} (January 1998).

\bibitem {Gue87}
  {\sc G{\"u}nther, C.: }\newblock {\em The polysymplectic {H}amiltonian
    formalism in field theory and calculus of variations. I: The local case.}
  \newblock J. Differ. Geom.  {\bf 25} (1987), 23--53.

\bibitem {Gue91}
  {\sc G{{\"u}}nther, C.: }\newblock {\em Polysymplectic quantum field theory.}
  \newblock In: {\sc Doebner, H.~D., Hennig, J.~D. (Eds.): }\newblock {\em
    Differential geometric methods in theoretical physics, 
    Proc. 15th Int. Conf., DGM, Clausthal/FRG, 1986}, 
  14--27, Singapore, 1987. World Scientific Publishing Co.

\bibitem {Kan97}
  {\sc Kanatchikov, I.~V.: }\newblock 
  {\em On field theoretic generalizations of a {P}oisson algebra.}
  \newblock Rep. on Math. Phys.  {\bf 40}.2 (1997), 225--234,
  \newblock {\tt hep-th/9710069}.

\bibitem {Kan98}
  {\sc Kanatchikov, I.~V.: }\newblock 
  {\em Canonical structure of classical field
    theory in the polymomentum phase space.}
  \newblock Rep. on Math. Phys.  {\bf 41}.1 (1998), 49--90,
  \newblock {\tt hep-th/9709229}.

\bibitem {Kan99}
  {\sc Kanatchikov, I.~V.: }\newblock {\em De {D}onder-{W}eyl theory and a
    hypercomplex extension of quantum mechanics to field theory.}
  \newblock Rep. on Math. Phys.  {\bf 43}.1--2 (1999), 157--170,
  \newblock {\tt hep-th/9810165}.

\bibitem {Kol87}
  {\sc Kol{\'a}{\v r}, I.: }\newblock {\em Some Natural Operations with
    Connections}.
  \newblock J. Nat. Acad. Math. India  {\bf 5}.2 (1987), 127--141.

\bibitem {KMS93}
  {\sc Kol{\'{a}}{\v{r}}, I., Michor, P.~W., Slov{\'{a}}k, J.: }
  \newblock {\em  Natural Operations in Differential Geometry}.
  \newblock Springer-Verlag, Berlin, Heidelberg, New York, 1993.

\bibitem {Noe18b}
  {\sc Noether, E.: }\newblock {\em Invarianten beliebiger
    {D}ifferentialausdr{\"u}cke.}
  \newblock Nachr. Kgl. Ges. Wiss. G{\"o}ttingen, Math.-phys. Kl. (1918), 37.

\bibitem {Ru66}
  {\sc Rund, H.: }\newblock {\em The {H}amilton-{J}acobi Theory 
    in the Calculus  of Variations}.
  \newblock Hazell, Watson and Viney Ltd., Aylesbury, Buckinghamshire, U.K.,
  1966.

\bibitem {Sau89}
  {\sc Saunders, D.~J.: }\newblock {\em The Geometry of Jet Bundles}.
  \newblock {\em Lond. Math. Soc. Lect. Note Ser., 142}.
  \newblock Cambr. Univ. Pr., Cambridge, 1989.

\bibitem {Trau67}
  {\sc Trautman, A.: }\newblock {\em Noether Equations and Conservation Laws}.
  \newblock Commun. Math. Phys.  {\bf 6} (1967), 248--261.

\end{thebibliography}
\end{document}